\documentclass[aps,pra,showpacs,twoside,twocolumn,10pt]{revtex4-1}
\usepackage[colorlinks=true, citecolor=red, urlcolor=blue ]{hyperref}
\usepackage{epsfig,newlfont,amssymb,amsfonts,amsmath,bm,subfigure,palatino,mathtools,amsthm,braket,times,soul}
\usepackage{color}
\usepackage{multirow}
\usepackage{ulem}

\definecolor{indiagreen}{rgb}{0.07, 0.53, 0.03}
	\definecolor{teal}{rgb}{0.0, 0.53, 0.53}

\begin{document}

\title{
Quantum illumination with a light absorbing target
	}

	\author{Rivu Gupta\(^1\), Saptarshi Roy\(^1\), Tamoghna Das\(^2\), and Aditi Sen(De)\(^{1}\)}
	
	\affiliation{\(^1\)Harish-Chandra Research Institute, HBNI, Chhatnag Road, Jhunsi, Allahabad 211 019, India}
	\affiliation{\(^2\)International Centre for Theory of Quantum Technologies,  University of Gda\'{n}sk, 80-952 Gda\'{n}sk, Poland.}

\begin{abstract}

In a quantum illumination (QI) protocol, the task is to detect the presence of the target which is typically modelled by a partially reflecting beam splitter. We analyze the performance of QI when the target absorbs part of the light that falls on it, thereby making the scenario more realistic. We present an optical setup that models a target with these characteristics and explore its detectability in the quantum domain in terms of the Chernoff bound (CB). For an idler-free setup, we use the coherent state for QI while the two mode squeezed vacuum (TMSV) state is employed in the signal-idler scheme. In both the cases, we report an \textit{absorption-induced enhancement}  of the detection efficiency indicated by a lowering of CB with increasing amounts of absorption. Interestingly, we show that in the presence of absorption, a more intense thermal background can lead to target detection with enhanced efficiency. Moreover, we observe that the  quantum advantage persists even for finite amounts of absorption. However, we find that the quantum advantage offered by TMSV decreases monotonically with absorption, and becomes vanishingly small in the high absorption regime. We also demonstrate the optimality of both the coherent and the TMSV states in their respective setups (idler-free and signal-idler) in the limit of low reflectivity and absorption. 

\end{abstract}
	
	\maketitle

\section{Introduction}

 Illumination \cite{Lloyd,TrepsArxiv,Illu22,Illu3,Illu4,Illu5,Illu6,Illu7} is a process by which  the presence  of an object immersed in a noisy background can be detected.  Interestingly, it was reported that quantum mechanical systems can lead to the enhancement in the performance of illumination over classical schemes, providing yet another avenue of quantum advantage   \cite{Lloyd,GIllu,GIllu4,Illu2}.
In quantum illumination (QI), the object to be detected (target) is modelled by a partially reflecting beam splitter, a multislab of dielectric plates, in a thermal background. If the target is present (hypothesis $H_1$), its existence is confirmed by analyzing the reflected light while in the absence of the target (hypothesis $H_0$), the detector merely gets the background thermal signature. Therefore, the problem of detecting the target  reduces to the distinguishability of  two quantum states,  $\rho_0$ and $\rho_1$ that  correspond to the hypotheses $H_0$ and $H_1$ respectively. The error probability, $p_e$, involved in distinguishing $\rho_0$ and $\rho_1$ quantifies the performance of the illumination process. A computable upper bound to the error is given by the quantum Chernoff bound (CB), $\mathcal{Q}$,  \cite{Chernoff,Chernoff2,Chernoff3,Bound1,Bhattacharyya},
\begin{eqnarray}
p_e \leq \frac{1}{2}\mathcal{Q} = \frac{1}{2} \min_{0\leq s \leq 1} \text{tr} (\rho_0^s \rho_1^{1-s}).
\end{eqnarray}
When $M$ copies of the probe are used, the above inequality modifies as 
$p_e(M) \leq \frac{1}{2}\mathcal{Q}^M$.  Moreover, the Chernoff bound is asymptotically tight \cite{Discriminate2,ChernoffTight}, i.e., the inequality saturates for large enough $M$. 
Therefore, to perform the entire analysis regarding  the performance of  QI,    
it is enough to compute the  Chernoff bound for a single copy to estimate the error incurred due to $M$ copies. Up to now, all the investigations on QI are concentrated on the choices of the probes, ranging from Gaussian  to non-Gaussian states in presence or in absence of noise \cite{GIllu6,GIllu7,GIllu8,AsymmSqueeze,GIllu10,NGIllu,OurIllu}.  

\begin{figure}
		\centering
		\includegraphics[width=\linewidth]{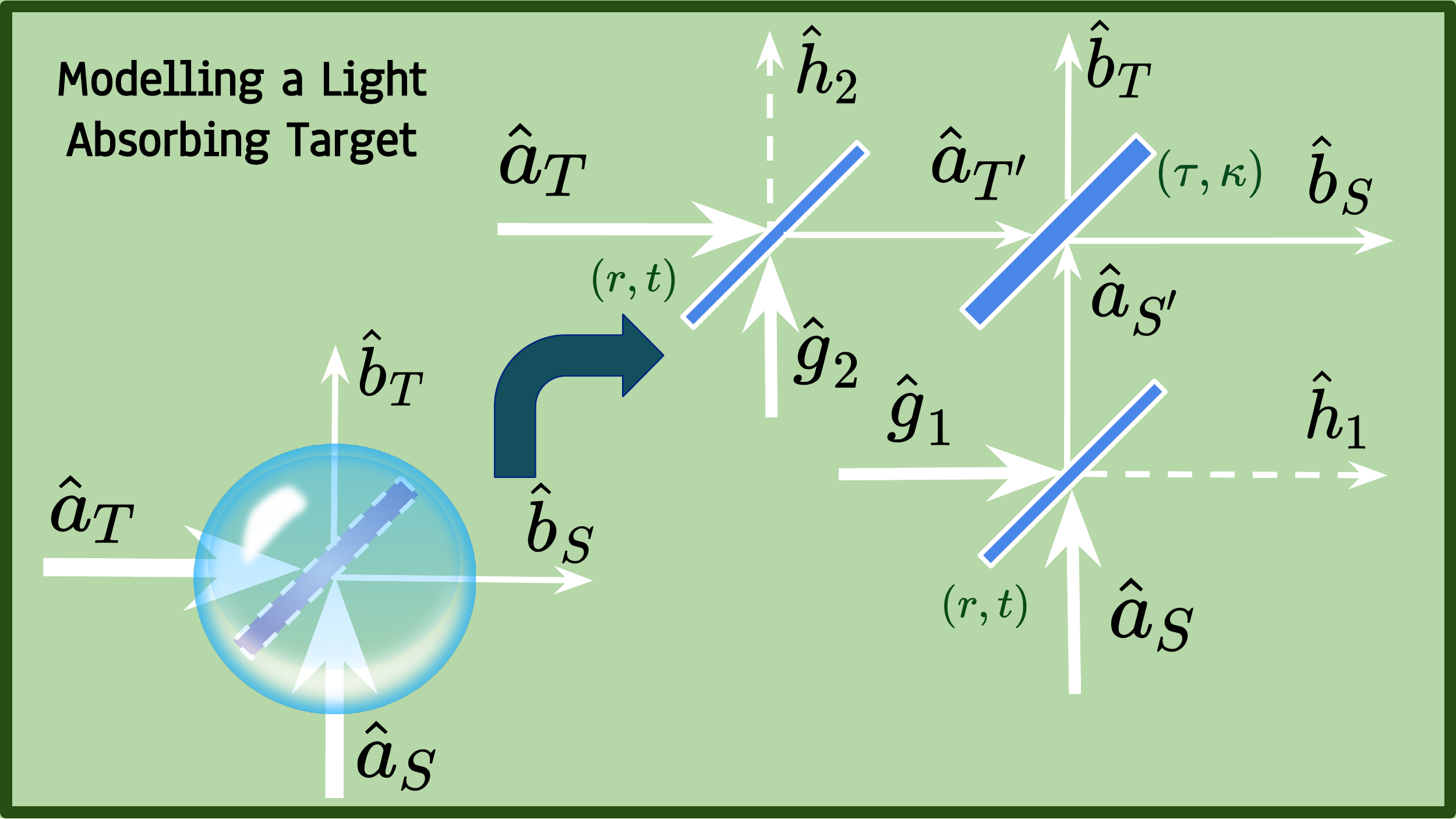}
		\caption{Schematic representation of a light absorbing target, modelled with three beam splitters, two auxiliary and one primary. The auxiliary beam splitters with reflectivity, $r$, and transmissivity, $t$, incorporates the loss in the signal and the thermal modes, while the primary beam splitter has reflectivity, $\kappa$, and transmissivity,  $\tau$, with $r, \,t,\, \kappa,\, \tau \geq 0$. The entire beam splitter setup constitutes the light absorbing target.}
		\label{fig:lossy_model}
	\end{figure}

In this paper, we introduce imperfection in a  different component involved in the illumination process. In particular, 
we conceive the target in a more realistic context where in addition to the  transmission and reflection, it absorbs some portion of the light impinged on it \cite{LossyScience,LossyAssym}. We model such a target using a lossy beam splitter (BS) \cite{Laudon_PRA,ImotoPRL,RMP1,RMP2}, naturally occurring in dielectric medium, where the loss quantifies the amount of absorption in the target. Note here that such an absorbing target which can be modelled by auxiliary  beam splitters \cite{lossbs} has already been used to assess quantum communication protocols like  quantum key distribution  \cite{LossyQKD}. 

 For introducing absorption in QI, we present an optical setup using three  lossless beam splitters, among them  two are the auxiliary beam splitters, modeling the absorption of the system while the rest one is the  primary beam splitter which  tunes the reflectivity and the transmissivity  of the same, as depicted in Fig. \ref{fig:lossy_model}. 
The two auxiliary beam splitters,  each of them having reflectivity $r$ and transmission coefficient, $t$, with $r + t = 1$,  incorporate the absorption in the optical modes, with the parameter $r$ being the absorption parameter dictating the amount of absorption. 
On the other hand, the partially transmitted signal and the thermal modes impinge on the primary beam splitter, having reflectivity and transmissivity, $\kappa$ and $\tau$ respectively. Note that the reflectivity of the absorbing target in terms of these parameters is $t \kappa$, while its transmissivity being $t \tau$. In the lossy scenario, we analytically compute  the Chernoff bounds both for  the coherent  as well as for the TMSV states and show that unlike the absorption-free situation, the Chernoff bounds for both the states do not reduce to the  Uhlmann's fidelity (the quantum version of the Bhattcharyya bound) \cite{Uhlmann76}.    We exhibit that the absorption of light leads to a lower probability of error both in case of the coherent and the TMSV state which we refer to as {\it absorption-induced enhancement}. Note that even in the presence of absorption, the TMSV state continues to outperform the coherent state. Moreover, we prove that the coherent state is an optimal idler-free probe in the limit of low absorption and target reflectivity, while in the signal-idler setup, the TMSV state turns out to be optimal under the same limiting conditions.


 The rest of the paper is organised in the following way. We first introduce a detailed formalism of the absorbing BS, highlighting the input-output relations of the optical modes in Sec. \ref{sec:model}. In Sec. \ref{sec:coherent}, we  proceed to calculate the analytical forms of the Chernoff bound for the coherent state and demonstrate its optimality in the idler-free illumination protocol.  We  analyse quantum illumination using the TMSV state in Sec. \ref{sec:TMSV} while  the quantum advantages of the illumination process with  a realistic target having different absorption parameters are shown in SubSec. \ref{subsec:qadv}. In Sec. \ref{sec:optimal}, we derive the optimal probes for quantum illumination of an absorbing target in the single mode and two mode regimes. Concluding remarks are presented in Sec. \ref{sec:conclu}. 

\section{Modelling absorption by a Lossy Beam Splitter} 
	\label{sec:model}
For a lossy beam splitter (LBS), if $\{b_1, b_2\}$ are the output optical modes and $\{a_1, a_2\}$ are the input optical modes, we have
\begin{eqnarray}
\langle b_1^{\dagger} b_1 + b_2^{\dagger} b_2 \rangle < \langle a_1^{\dagger} a_1 + a_2^{\dagger} a_2 \rangle,  \end{eqnarray}
which directly follows from the condition of the lossy beam splitter \cite{Laudon_PRA},  given by
		\begin{equation}
		\mathcal{T} + \mathcal{R} < 1,
		\end{equation}
where $\mathcal{T}$ and $\mathcal{R}$ respectively denote the transmission and reflection coefficients of the LBS.
It means that if $N$ number of photons impinge on both the input modes of the lossy beam splitter,  the total number of photons at the output modes is strictly less than $ N$, implying that some other modes intrinsic to the system (beam splitter) get excited.
Let $\{g_1, g_2\}$  and $\{h_1, h_2\}$ be the device input modes and  output modes respectively. All the output modes $\{\hat{\mathbf{b}}, \hat{\mathbf{h}}\}$ (optical + device) are related to  all  the input modes $\{\hat{\mathbf{a}}, \hat{\mathbf{g}}\}$ by the matrix $\hat{\Lambda}$  \cite{Welsh1} such that	
		\begin{eqnarray}
		\begin{pmatrix}
		\hat{\mathbf{b}} \\ \hat{\mathbf{h}} 
		\end{pmatrix} = \begin{pmatrix}
			\mathrm{\hat T} & \mathrm{\hat A} \\
			\mathrm{\hat F} & \mathrm{\hat G}
			\end{pmatrix} . \begin{pmatrix}
		\hat{\mathbf{a}} \\ \hat{\mathbf{g}} 
		\end{pmatrix} = \Lambda . \begin{pmatrix}
		\hat{\mathbf{a}} \\ \hat{\mathbf{g}} 
		\end{pmatrix}.
		\end{eqnarray}
The output optical modes are given by
$\hat{\mathbf{b}} = \mathrm{\hat T}\hat{\mathbf{a}} + \mathrm{\hat A}\hat{\mathbf{g}},$
where $\mathrm{\hat T}$ and $\mathrm{\hat A}$ denote the transmission and absorption matrices respectively.
Imposing the unitarity condition which in turn implies the photon number conservation (including the photons that get absorbed), we have
\begin{eqnarray}
\mathrm{\hat T}^{\dagger}\mathrm{\hat T} + \mathrm{\hat A}^{\dagger}\mathrm{\hat A} = \mathbb{I},
\end{eqnarray}
with \(\mathbb{I}\) being the identity operator. 
Note that $\mathrm{\hat A}^{\dagger}\mathrm{\hat A} > 0$ refers to a finite absorption. In our model, as depicted in Fig. \ref{fig:lossy_model}, the transmission and absorption matrices are taken respectively as
\begin{eqnarray}
\mathrm{\hat T}^{\dagger}\mathrm{\hat T} = t ~\mathbb{I}, \text{ and } \mathrm{\hat A}^{\dagger}\mathrm{\hat A} = (1-t) ~\mathbb{I}.
\end{eqnarray}
Naturally,  $1-t = r$, i.e., the reflectivity of the auxiliary BS,  which controls the amount of absorption. For a detailed derivation of the $\mathrm{\hat T}$ and $\mathrm{\hat A}$ matrices, see Appendix \ref{app:1}. 
Our approach of modelling absorption process using two auxiliary beam splitters  has an added advantage since it provides an actual optical setup that replicates the loss, giving a physical interpretation to the device modes as well.  


\section{target detection in an idler-free setup} 
\label{sec:coherent}

When the target absorbs some of the light that impinges on it, we find that its presence can be detected more efficiently compared to the absorption-free case.  We will show in this section and in the succeeding sections that the enhancement features persist  both for the classical (idler-free) and quantum illumination schemes --  {\it absorption-induced enhancement}.
Here, we  sketch the derivation of the Chernoff bound for the coherent state. 


\subsection{Chernoff bound for the coherent state}
\begin{figure}
    \centering
    \includegraphics[width = \linewidth]{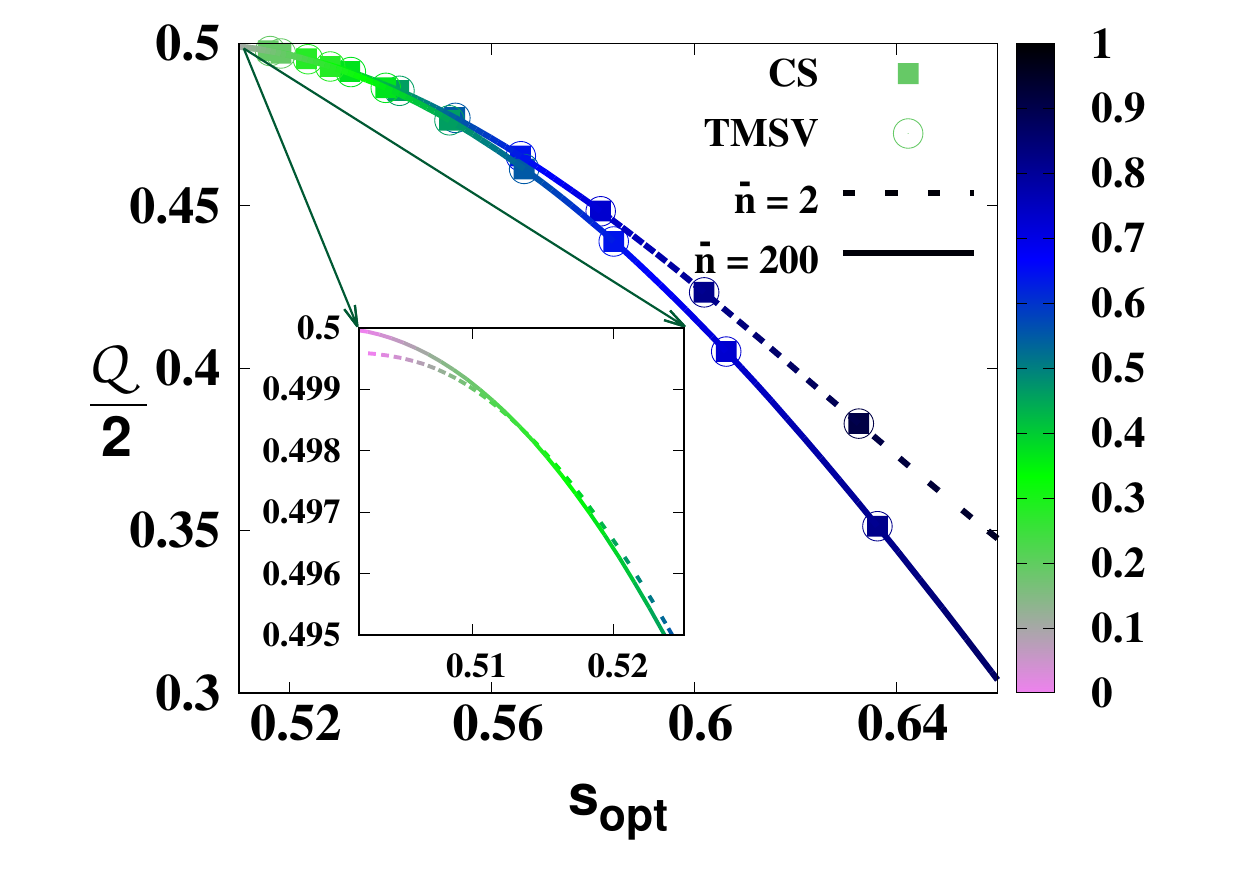}
    \caption{The upper bound for the error probability i.e. $\frac{1}{2} \mathcal{Q}$ (ordinate) against the optimal value of $s$, denoted by \(s_{opt}\) (abscissa). The color scheme (light to dark) denotes different absorption parameters of the target, and we show that $s_{opt}$ decreases with an increase in the amount of light absorption. The solid squares represent the coherent state while  the points for the TMSV state  are the hollow circles. The solid lines indicate a mean thermal background strength of $\bar{n} = 2$ and the dotted lines depict $\bar{n} = 200$. The inset shows a magnified version of a region at very low absorption, where there is a crossover between the plots corresponding to low and high background thermal noise. For vanishing absorption, $\bar{n} = 200$ gives a higher error probability, but as the absorption increases, higher background noise leads to a lower value of CB, $\frac{1}{2}\mathcal{Q}$.  Here, the signal strength, $N_S = 0.5$, and we set $\kappa = 0.01$. Both the axes are dimensionless. }
    \label{fig:fig2}
\end{figure}

Here, the objective is to evaluate  the distance between $\rho_0$ (when the target is absent) and the resulting state, $\rho_1$ after the coherent state (CS) of signal strength $N_S$, $|\psi \rangle_S = \exp({-N_S/2}) \sum_{n = 0}^{\infty} (N_{S}^{n}/n!)^{1/2}|n \rangle$ passes through the optical setup that models  the light-absorbing target as depicted in Fig. \ref{fig:lossy_model}.
  When the target is absent, $\rho_0 = \rho_T = \sum_{m = 0}^{\infty} \bar{n}^m/(1 + \bar{n})^{m+1} |m \rangle \langle m |$, which represents the thermal background having mean photon number $\bar{n}$. As the combined actions of the beam splitters in Fig. \ref{fig:lossy_model} preserve Gaussianity, we can analytically compute $\rho_1$ in the phase space using symplectic operations corresponding  to the BS setup in terms of the displacement vector and the covariance matrix (for detailed calculations, see Appendix. \ref{app:2}). The displacement vector and the covariance matrix in the target-absent ($\rho_0$) and -present ($\rho_1$) scenarios have the following expressions:
\begin{eqnarray}
&& \rho_0: \mathbf{d}_0 = \begin{pmatrix}
0 \\
0
\end{pmatrix}, ~~~~~~~~~~~~~~ \mathbf{\sigma}_0 = (2 \bar{n} + 1) \mathbf{I}_2;
\label{coh_rho0_dS}
 \\ \nonumber
\\ 
&&  \rho_1: \mathbf{d}_1 = \begin{pmatrix}
2 \sqrt{N_S \kappa t} \\
0
\end{pmatrix}, ~~~ \mathbf{\sigma}_1 = (2 \bar{n} \tau t + 1) \mathbf{I}_2.
\label{coh_rho1_dS}
\end{eqnarray}
The subscripts, \(i\), of the displacement and the covariance matrix denote the corresponding \(\rho_i\)s (\(i=0,1\)) and $\mathbf{I}_2 = \text{diag}\{1,1\}$. We will now denote the states in their normal mode form as $\rho_i = (\boldsymbol{d_i}, \boldsymbol{\mathbb{S}_i},\{\nu_{i}^k\})$, where $\boldsymbol{d_i}$ is the displacement vector and  $\boldsymbol{\mathbb{S}_i}$ is the symplectic matrix having a set of $k$ symplectic eigenvalues $\{\nu_{i}^k\}$ \cite{Bound1}.
In this case, the symplectic decompositions take the forms as
\begin{eqnarray}
&& \nonumber \rho_0 \equiv (\boldsymbol{d_0}, \mathbf{I_2}, {2 \bar{n} + 1}), ~ \text{and} ~ \\ &&
 \rho_1 \equiv (\boldsymbol{d_1}, \mathbf{I_2}, {2 \bar{n}t\tau + 1}). ~~~
\label{coh_symp}
 \end{eqnarray}
Using Eq. \eqref{coh_symp} and following the prescription given in Ref. \cite{Bound1},  the Chernoff bound for the coherent state reads as $\mathcal{Q}_{CS} = \min_{0 \leq s \leq 1} [\mathcal{Q}_s^{CS}]$, where
\begin{equation}
    \mathcal{Q}_{s}^{CS} =  \frac{\exp\Big(-\frac{\kappa t N_S}{\frac{\bar{n}^s}{(1 + \bar{n})^s - \bar{n}^s} + (1 - (\frac{\bar{n}\tau t}{1 + \bar{n}\tau t})^{1-s})^{-1}}\Big)}{(1 + \bar{n})^s  (1 + \bar{n} \tau t)^{1-s} - \bar{n} (\tau t)^{1-s}}.
    \label{coh_CB}
\end{equation}
Unlike the usual illumination scheme with a non-absorbing target, we interestingly find that the parameter $s$ in this lossy scenario does not have a unique value $(=\frac{1}{2})$ in Eq. \eqref{coh_CB}, i.e., the value of \(s\) can be determined depending on the amount of absorption and the background photon strength.
For a fixed $\bar{n}$, the optimal $s$ increases with the absorption parameter $r$ as shown in Fig. \ref{fig:fig2}. Notice that it only reduces to the Uhlmann's fidelity \cite{Bhattacharyya}, i.e., with \(s=1/2\) in the limiting case. 

We observe another striking feature which is completely different from the absorption-free case, where, as expected, the detection efficiency decreases when the background thermal noise increases. However, when the absorption parameter is more than a critical value (\(\gtrsim 0.1\)) (see inset of Fig. \ref{fig:fig2}), a higher $\bar{n}$ actually leads to a lower Chernoff bound. Although this feature of noise-induced enhancement of target detection at the first glance seems counter-intuitive, we can provide a physical explanation for this. In  presence of absorption and low reflectivity, the light that reaches the detector is only a fraction of the signal intensity combined with background noise (which also suffers absorption, a feature which is absent in a non-absorbing target) $(\rho_1)$, while  in the absence of the target, a strong thermal background  implies that the detector senses a high-intensity light $(\rho_0)$. It makes the  photon number difference between $\rho_0$ and $\rho_1$ quite high resulting in a more efficient distinguishability.

\section{Absorption improves target detection --  Analyzing quantum advantage}
\label{sec:TMSV}

In this section, we focus on the quantum advantages obtained in the QI via the signal-idler setup using the two mode squeezed vacuum (TMSV) state, over the coherent state-based optimal classical scheme, in the lossy scenario. We first compute $\mathcal{Q}$ for the TMSV state in this domain and    demonstrate a monotonic enhancement of the quantum advantage with respect to the absorption parameter of the target which we illustrate by considering two regimes -- intermediate domain and limiting cases.

The two mode squeezed vacuum state, given by $ |\psi \rangle_{TMSV} = \sum_{n  = 0}^{\infty} \sqrt{N_S^n}/\sqrt{(1 + N_S)^{1+n}} |n \rangle |n \rangle$,  is  shown to be the optimal two mode probe state for detecting a non-absorbing target \cite{Optimalprobes}. It was also proven that it can outperform the coherent state-based illumination protocol  in the low signal strength and high background noise limit \cite{GIllu}, thereby establishing the quantum advantage.

 Here, we aim to derive an expression for the Chernoff bound achieved by the TMSV state, when trying to detect an absorbing target. For a TMSV state of mean signal strength $N_S$, the displacement vector and the covariance matrix in the absence ($\rho_0$) and presence ($\rho_1$) of an absorbing target have the following  expressions:
\begin{eqnarray}
\boldsymbol{d_0} = \boldsymbol{d_1} = (0 ~~~ 0 ~~~ 0 ~~~ 0 ~)^T, \label{rho_disp_t}\\
\boldsymbol{\sigma_0} = \begin{pmatrix}
B_0 & 0 & 0 & 0\\
0 & B_0 & 0 & 0 \\
0 & 0 & S & 0 \\
0 & 0 & 0 & S
\end{pmatrix}, 
\label{cov_rho_0_t} \\
\boldsymbol{\sigma_1} = \begin{pmatrix}
A \mathbf{I_2} & \sqrt{\kappa} C_q \mathbf{Z_2} \\
\sqrt{\kappa} C_q \mathbf{Z_2} & S \mathbf{I_2}
\end{pmatrix},
\label{cov_rh01_t}
\end{eqnarray}
where $S = 2N_S + 1, B_0 = 2\bar{n} + 1, B = 2\bar{n}t \tau + 1, A = 2\kappa N_S t + B$ and $C_q = 2\sqrt{tN_S(N_S+1)}$. The normal mode decompositions of $\rho_0$ and $\rho_1$ thus become \cite{GIllu}
\begin{eqnarray}
&&\rho_0 \equiv (\boldsymbol{d_0}, \mathbf{I_4}, {B_0, S}), \,\, \mbox{and} \label{rho0t_symp} \\
&& \rho_1 \equiv (\boldsymbol{d_1}, \boldsymbol{\mathbb{S}_1}, {\nu_1, \nu_2} ), \label{rho1t_symp}
\end{eqnarray}
where $\nu_k = \frac 12 \left((-1)^{k}(S-A)+\sqrt{(A+S)^{2}-4 \kappa C_{q}^{2}}\right) $ with $\mathbf{I}_4 = \text{diag} \{1,1,1,1\}$. The symplectic $\boldsymbol{\mathbb{S}_1}$ which diagonalises $\boldsymbol{\sigma_1}$, reads as
\begin{equation}
    \boldsymbol{\mathbb{S}_1} = \begin{pmatrix}
    x_+ \mathbf{I_2} & x_- \mathbf{Z_2} \\
    x_- \mathbf{Z_2} & x_+ \mathbf{I_2}
    \end{pmatrix},
    \label{symp_tmsv}
\end{equation}\\
where $x_{\pm} = \sqrt{\frac{A+S \pm \sqrt{(A+S)^{2}-4 \kappa C_{q}^{2}}}{2 \sqrt{(A+S)^{2}-4 \kappa C_{q}^{2}}}}$ and $\mathbf{Z}_2 = \text{diag} \{ 1,-1\}$.
The analytical form of CB for the TMSV state, is $\mathcal{Q}_{TMSV} = \min_{0 \leq s \leq 1}[\mathcal{Q}_s^{TMSV}]$, where
\begin{equation}
    \mathcal{Q}_s^{TMSV} = \Big[ \frac{4 G_s(B_0) G_s(S) G_{1-s}(\nu_1) G_{1-s}(\nu_2) }{ \left(\Sigma_{1-s}^+ + \Lambda_s(B_0)\right)\left( \Sigma_{1-s}^- + \Lambda_s(S) \right) - \Omega_{1-s}^2}\Big],
\end{equation}
where $\Sigma_{1-s}^+ = \left(\Lambda _{1-s}(\nu _1) x_+^2 + \Lambda _{1-s}(\nu _2) x_-^2\right)$, $\Sigma_{1-s}^- = \left(\Lambda _{1-s}(\nu _1) x_-^2 + \Lambda _{1-s}(\nu _2) x_+^2\right)$ and $\Omega_{1-s} = \left(\Lambda _{1-s}(\nu _1) + \Lambda _{1-s}(\nu _2)\right) x_- x_+$.
The function  $G_p(x)$ and $\Lambda_p(x)$ are respectively given by \cite{Bound1}
\begin{eqnarray}
    G_{p}(x) = \frac{2^{p}}{(x+1)^{p}-(x-1)^{p}}, \\
    \Lambda_{p}(x) = \frac{(x+1)^{p}+(x-1)^{p}}{(x+1)^{p}-(x-1)^{p}}.
    \label{lambda_s}
\end{eqnarray}
However, one can find that the value of $s$ which minimises $Q_s$ to furnish the Chernoff bound again depends on both the loss parameter $r$ and the mean background noise strength $\bar{n}$.

\noindent \textit{Remark.} Up to numerical accuracy, $s_{opt}$ for the TMSV and the coherent states are almost identical for the same set of system parameters, i.e., their difference is $O(10^{-4})$ or less.

\begin{figure*}
    \centering
    \includegraphics[width = \linewidth]{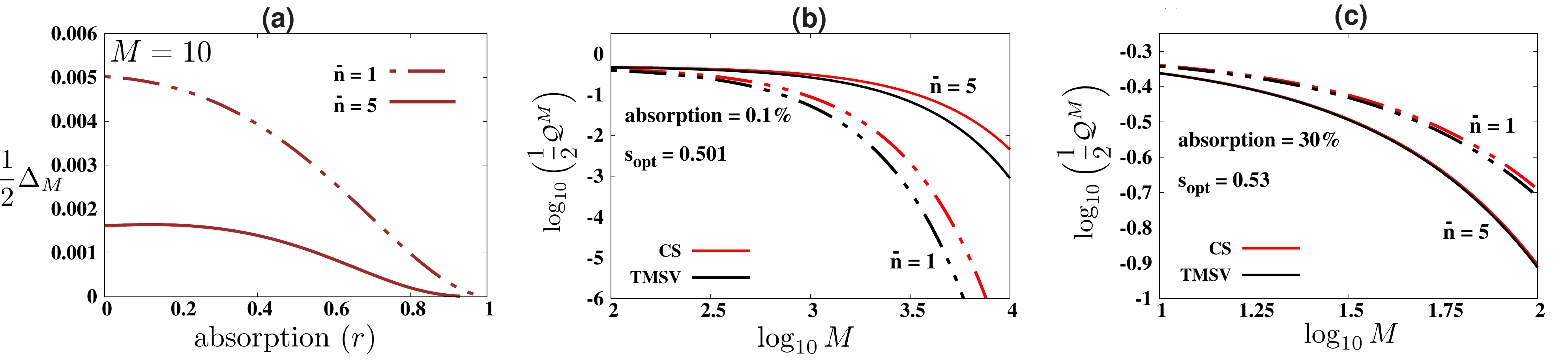}
    \caption{The variation of the quantum advantage ($\frac{1}{2}\Delta_M$) and the Chernoff bound  against different parameters for the coherent and the TMSV states having mean signal strength $N_S = 1.0$  and  $\kappa = 0.01$. The dotted lines represent a background noise having $\bar{n} = 1$ and the solid lines correspond to $\bar{n} = 5$. (a) Quantum advantage, $\frac{1}{2}\Delta_{M = 10}$  (ordinate), with respect tot the absorption parameter, $r$, (abscissa) for varying background thermal noise. It clearly demonstrates the decrease in quantum advantage with increasing absorption parameter. (b) Logarithm of $\frac{1}{2}\mathcal{Q}^M$ (\(y\)-axis) vs. logarithm of the number of copies ($\log_{10} M$) (\(x\)-axis) in the limit of the low absorption parameter ($0.1\%$) for the coherent state (in red) and the TMSV state (in black). It shows that the error probability worsens with an increase of the background noise strength. (c) Same plot as panel (b)  for the intermediate absorption regime ($30 \%$). In this case, increase in thermal noise assists in target detection. All the axes are dimensionless. }
    \label{fig:fig3}
\end{figure*}

\subsection{Chernoff bound vs absorption: Quantum advantage of TMSV}
\label{subsec:qadv}

In the previous section, we have derived the quantum Chernoff bound of the TMSV state, for a light absorbing target, having reflectivity $t\kappa$, and absorption coefficient $r$. 
Here, we will discuss about the same for the different limit of absorption coefficients.
Depending on the amount of absorption, one can identify three regimes --   low, intermediate and high absorption parameters and we discuss some novel features of each regime. On the other hand, Eq. \eqref{coh_CB} can represent the error probability in detecting an absorbing target using the coherent state as the probe and hence comparing the quantities for CS and the TMSV state, we can establish the quantum advantage in the illumination process with light absorbing target provided the CS leads to an optimal classical probe which we will discuss in the next section. 

\subsubsection{Intermediate regime}

When the absorbing power of the target is neither too high nor too low, the exact Chernoff bound  depends on the optimization of $s$ which is again dependent on the absorption parameter $r$ as well as the mean background photon number $\bar{n}$ and hence cannot be analytically obtained. We  numerically verify that when $s \in (0.5,1)$, the Chernoff bound decreases with the increasing signal strength $N_S$. The counter-intuitive aspect of the absorbing target is that the detection is enhanced by the presence of absorption. As the target becomes more absorbing, the Chernoff bound goes down steadily, as shown in Fig. \ref{fig:fig2}. It may be explained due to the loss of background thermal noise, which too is absorbed by the target. In other words, in the event of no target, the thermal state reaches the detector while  in the presence of the target, the state at the detector  is a mixture of a highly absorbed signal and thermal states. Hence distinguishability is enhanced as compared to the case of the non-absorbing target, and the  error probability in detection goes down. These results clearly demonstrate the {\it constructive effect of an absorbing target} towards its detection. The higher the absorbing power, the more easily can the reflected signal be analysed to infer its presence. In fact, for absorption-less targets, the presence is assessed by analyzing the reflected (presence of) light. On the contrary, for light absorbing targets, their presence is inferred from both the reflected (presence of) light and the absorbed (absence of) light. The absorbing nature of the target yields a better detection probability due to the fact that the background thermal noise is also equally affected. Thus, for a light absorbing target, the background thermal noise enhances detection efficiency as the absorbing power of the target increases (see Fig. \ref{fig:fig3}(c)).

Let us now address the question whether the absorption-induced enhancement in target detection  translates to quantum advantage, which for $M$ copies reads as
\begin{eqnarray}
\Delta_M = \mathcal{Q}_{CS}^M - \mathcal{Q}_{TMSV}^M.
\label{eq:qa}
\end{eqnarray}
Fig. \ref{fig:fig3} (a) depicts that the quantum advantage persists even in presence of absorption although \(\Delta_M\) decreases with the increase of \(r\). 
The trends of \(\Delta_M\) indicates that  the performance of the TMSV and coherent states as probes  approach each other with increasing absorbing power. Therefore, an absorbing target leads  to a detection with a much lower error probability and also makes the classical protocol as good as compared to the TMSV state-based quantum illumination scheme for high absorption.

\subsubsection{Limiting cases: low and high absorption}

When the absorbing power of the target is low i.e., in the limit $r \to 0$, it can be shown that $s \to \frac 12$ as depicted in Fig. \ref{fig:fig2},   and the Chernoff bound for the coherent state, with $\kappa r \to 0$ and $\bar{n} >> 1$, reduces to
\begin{equation}
    \mathcal{Q}_{CS} = \exp ~\Big[-\frac{ \kappa N_S}{2\bar{n}(2-r-\kappa) } \Big].
    \label{low_loss_coh_CB}
\end{equation}
In the limit of low $N_S$, we provide the following ansatz for the Chernoff bound of the TMSV state: 
\begin{equation}
    \mathcal{Q}_{TMSV} = \exp ~\Big[-\frac{\kappa N_S}{\bar{n}(1 - r - \kappa)} \Big],
    \label{low_loss_coh_TMSV}
\end{equation}
which matches with numerical calculations upto an accuracy of $10^{-5}.$ Clearly, for  a given signal strength $N_S$, we see that $\mathcal{Q}_{TMSV} < \mathcal{Q}_{CS}$, thereby guaranteeing quantum advantage in the low loss regime (see Fig. \ref{fig:fig3}(b)). In this limit, a higher background noise hampers the detection probability, as in the case of a non-absorbing target.
Furthermore, it is worthwhile to mention that the quantum advantage is more in the case of a very low absorption regime compared to the completely absorption-free  \((r=0)\) scenario.


Let us now move to another extreme situation, i.e., in the limit $r \to 1$ when the target approaches the perfect absorption regime. In this scenario, both the signal and the thermal states are almost completely absorbed by the target. 
It implies that  no light reaches the target and hence its presence has to be assessed by the absence of light. 
Thus, for target detection, the state discrimination problem reduces in differentiating between $\rho_0 = \rho_T$ and $\rho_1 = |0\rangle\langle0|$ which we obtain by substituting $r \to 1$ (or equivalently $t \to 0$) in Eq. \eqref{coh_rho1_dS} 
for which the Chernoff bound can be expressed as
  \( \mathcal{Q}_{CS}= \frac{1}{1 + \bar{n}}\).
Interestingly, the Chernoff bound turns out to be equal to the probability of the vacuum mode being present in the thermal mode for the coherent state.
In case of probing with the TMSV state, the state discrimination problem is between the states $\rho_0 = \rho_T \otimes \rho_I$ and $\rho_1 = |0\rangle\langle0|\otimes \rho_I$, where $\rho_I$ is the reduced state obtained by tracing out the signal mode in $|\psi\rangle_{TMSV}$ which turns out to be a thermal state with mean photon number $N_S$. Since $\rho_I$ is a fixed addition in both $\rho_0$ and $\rho_1$, it does not aid in enhancing the detection probability, and we get
 \(\mathcal{Q}_{TMSV} = \mathcal{Q}_{CS} =  \frac{1}{1 + \bar{n}}\). 
Some important features emerge from our analysis. Firstly, in both the cases,  the optimal $s$ parameter in the Chernoff bound approaches unity to minimize it. Secondly, the Chernoff bound decreases with the background thermal noise strength $\bar{n}$, implying a better detection for higher noise values with $\mathcal{Q} \to 0$ for $\bar{n} \to \infty$. Finally, the quantum advantage, as defined in Eq. \eqref{eq:qa}, 
 identically vanishes irrespective of $\bar{n}$ and $N_S$ for $r \to 1$.

\section{Optimal probes for  detection of light absorbing targets} 
\label{sec:optimal}

In this section, we show that, in the limit of low absorbing power and low reflectivity of the target, the coherent state provides the optimal illumination scheme in the single mode regime, whereas the same in the two mode case is exhibited by the TMSV state. We base our analysis on the multiple beam splitter model as depicted in Fig. \ref{fig:lossy_model}. When the low reflectivity condition of the target is considered in terms of the primary BS, it can be approximated as
\begin{equation}
    U(\kappa) = \text{exp}[\theta (\hat{a}^\dagger \hat{b} - \hat{a}\hat{b}^\dagger)] \approx 1 + \sqrt{\kappa}(\hat{a}^\dagger \hat{b} - \hat{a}\hat{b}^\dagger),
    \label{BS_approx_primary}
\end{equation}
where the reflectivity parameter $\sqrt{\kappa} = \sin \theta \approx \theta$ in the low reflectivity approximation which leads to a significant contribution in the Chernoff bound as we will show later. On the other hand, in the low loss regime, the auxiliary beam splitters  having their reflectivity parameter, satisfying the condition, $\sqrt{r} = \sin \theta' \approx \theta'$  are approximated as
\begin{equation}
    U(r) = \text{exp}[\theta' (\hat{a}^\dagger \hat{b} - \hat{a}\hat{b}^\dagger)] \approx 1 + \sqrt{r}(\hat{a}^\dagger \hat{b} - \hat{a}\hat{b}^\dagger) + \frac{r}{2}(\hat{a}^\dagger \hat{b} - \hat{a}\hat{b}^\dagger)^2.
    \label{BS_approx_aux}
\end{equation}
The auxiliary BSs are approximated upto the second order of their reflectivity parameter, since we will see that the contribution to the Chernoff bound due to the absorption is absent for lower order terms. 

To calculate the minimum error probability in detecting a light absorbing target in the limit of low reflectivity and vanishing loss, the detected state  in case of the target being present  can be written by expanding about $\rho_0$ as
\begin{equation}
\rho_1 = \rho_0 + \epsilon (\sqrt{\kappa} \delta \rho_a + \sqrt{r} \delta \rho'_b + r \delta \rho_b + \sqrt{r \kappa} \delta \rho_c),
\label{perturbation_rho1}
\end{equation}
where $\kappa \to 0$ is the low reflectivity and $r \to 0$ represents the low loss approximation.  $\delta \rho_a$ is given by $\partial \rho_1/ \partial \sqrt{\kappa}|_{\kappa \to 0}$ while $\delta \rho'_b = \partial \rho_1/ \partial \sqrt{r}|_{r \to 0}$, $\delta \rho_b = \frac{1}{2}\partial \rho_1/ \partial r|_{r \to 0}$ and $\delta \rho_c = \partial \rho_1/ \partial \sqrt{r \kappa}|_{\kappa, r \to 0}$  \cite{Optimalprobes}. At the end of our calculations, we take $\epsilon \to 1$.

In this case, the Chernoff bound can be expressed as \cite{Discriminate2} $\mathcal{Q} = 1 - \Xi$, where
\begin{equation}
\Xi = \frac{\epsilon^2}{2} \sum_{j,k} \frac{|\langle j | ( \sqrt{\kappa} \delta \rho_a + \sqrt{r} \delta \rho'_b + r \delta \rho_b + \sqrt{r \kappa} \delta \rho_c) | k \rangle |^2}{(\sqrt{\chi_j} + \sqrt{\chi_k})^2}.
\label{CB_delta}
\end{equation}
Here $|i \rangle$ are eigenvectors of $\rho_0$ with corresponding eigenvalues $\chi_i$. The background thermal noise is considered to be the thermal state $\rho_{th} = \sum_{m = 0}^\infty \lambda_m |m \rangle \langle m |$ where $\lambda_m = \frac{\bar{n}^m}{(1 + \bar{n})^{m+1}}$, with $\bar{n}$ denoting the mean photon number.

\subsection{Single mode optimal probes}
 In case of a lossless target, the single mode optimal probe is known to be the coherent state \cite{Illu2,Optimalprobes}. Let us now identify the optimal probes for detecting a target modelled by a lossy beam splitter.  We demonstrate that, even if the target is an absorbing one, the coherent state still remains optimal as probes in the single mode case.\\

\noindent \textbf{Theorem 1}. A single mode state with mean photon number $N_S$, which provides the optimal Chernoff bound in illumination for a light absorbing target, is the coherent state when the reflectivity of the primary beam splitter is low in the limit of low absorption and low reflectivity. \\
\textit{Proof}. A single mode state in the Fock basis can be represented as
\begin{equation}
|\psi \rangle = \sum_{n = 0}^\infty C_n |n \rangle,
\label{single_mode}
\end{equation}
 and in this case, $\rho_1$ takes the  form as
\begin{eqnarray}
\nonumber \rho_1 = \sum_{m=0}^\infty && \sqrt{\kappa} \sqrt{(m+1)} (\lambda_{m+1} - \lambda_{m}) \times \\ \nonumber && ( \zeta|m \rangle \langle m+1 | + \zeta^*|m+1 \rangle \langle m|) +\\
&& ( (1 - r m) \lambda_m + r (m+1) \lambda_{m+1} ) |m \rangle \langle m| ~~~~~
\end{eqnarray}
where $\zeta = \sum_n C_nC^*_{n+1} \sqrt{n+1}$. Clearly, $\delta \rho'_b = \delta \rho_c =  0$, which means that the leading order contribution due to absorption comes from the quadratic terms of the reflectivity parameter of the auxiliary beam splitters. The states $\delta \rho_a$ and $\delta \rho_b$ in Eq. \eqref{perturbation_rho1} are given by
\begin{eqnarray}
\delta \rho_a = && \nonumber \sum_m \sqrt{m+1} (\lambda_{m+1} - \lambda_{m}) \times \\ && (\zeta |m \rangle \langle m+1| + \zeta |m+1 \rangle \langle m |), \label{delta_rho1}\\
 \delta \rho_b = && \frac{1}{2}\sum_{m}  (\lambda_{m+1} (m+1) - m \lambda_m  ) |m \rangle \langle m|. \label{delta_rho2}
\end{eqnarray}
Therefore, $\mathcal{Q}$ in Eq. \eqref{CB_delta} reduces to
\begin{equation}
\mathcal{Q} = 1 - \frac{\epsilon^2}{2} \left[\sum_{\substack{j \neq k\\|j-k| = 1}} \frac{|\langle j| \sqrt{\kappa} \delta \rho_a | k \rangle |^2}{(\sqrt{\chi_j} + \sqrt{\chi_k})^2}  + \sum_{j} \frac{|\langle j| r \delta \rho_b | j \rangle |^2}{4 \chi_j} \right]. ~~~~~~~~~~
\label{Q_coherent}
\end{equation}
In this case, $\{|j \rangle\}$, and $\{|k \rangle\}$ correspond to the eigenvectors $|m\rangle$ of the thermal state and $\chi_i = \lambda_i$, since $\rho_0 = \rho_{th}$ in case of the single mode probe.
If both the coefficients of \(\epsilon^2\) in  Eq. (\ref{Q_coherent}) are maximized by the coherent state, we can conclude that it is the optimal probe for a single mode illumination protocol. Firstly, the second subtracted term is the same for all single mode states, since it is independent of any state parameter and depends only on the background noise. Thus the minimization of $\mathcal{Q}$ depends only on the first subtracted term, which is given by
\begin{equation}
\kappa \sum_{m,n = 0}^\infty \frac{(m+1) (\lambda_m - \lambda_{m+1})^2 (|\zeta|^2 + |\zeta^*|^2)}{(\sqrt{\lambda_m} + \sqrt{\lambda_{m+1}})^2},
\label{firstQ}
\end{equation}
where $\zeta$ contains the dependence on the probe state. It is evident that the optimisation of Eq. (\ref{Q_coherent}) depends only on the magnitude of $\zeta$. In Ref. \cite{Optimalprobes}, it has been shown that since $\zeta = \sum_n C_n C^*_{n+1} \sqrt{n+1} \leq \sum_n |C_n||C_{n+1}| \sqrt{n+1}$, we can consider the $C_n$s to be real. Thus, subject to the normalisation ($\sum_n C_n^2 = 1$) and the energy ($\sum_n n C_n^2 = N_S$) constraints, the maximization condition for $\zeta$ reduces to \cite{Optimalprobes}
\begin{equation}
C_{n+1} \sqrt{n+1} + C_{n-1} \sqrt{n} + 2C_n (\mu_1 + n \mu_2) = 0
\end{equation}\\
which is satisfied by a coherent state of signal strength $N_S$ for Lagrange multipliers $\mu_1 = -\frac{\sqrt{N_S}}{2}$ and $\mu_2 = -\frac{1}{2\sqrt{N_S}}$. Therefore, the coherent state maximises the subtracted terms in Eq. \eqref{Q_coherent} and therefore minimises the Chernoff bound. Thus, a choice for the optimal single mode state for illumination of a light absorbing target is a coherent state with mean signal strength $N_S$. $\blacksquare$

\subsection{Two mode optimal probes}

Let us now move to the scenario of quantum illumination process where two mode states are considered as probes. It was shown that for a low reflectivity, TMSV state act as an optimal probe in the no-absorbing domain \cite{Illu2,Optimalprobes}. We will now show that the TMSV state still remains optimal in the absorbing  regime. \\
\textbf{Theorem 2}. The two mode state providing the optimal illumination protocol for a light absorbing target, is the TMSV state with mean signal photon number $N_S$ in the limit of low reflectivity and low absorption by the target. \\
\textit{Proof}. An optimal two mode probe state can be represented as \cite{Optimalprobes}
\begin{equation}
    |\psi \rangle = \sum_{n = 0}^\infty C_n |n, n \rangle,
\end{equation}
with
\begin{eqnarray}
    \delta \rho_a && = \nonumber \sum_{n,m} C_n C_{n+1} \sqrt{(n+1)(m+1)} (\lambda_{m+1} - \lambda_{m}) \times \\ && (|n,m \rangle \langle n+1,m+1| + |n+1,m+1 \rangle \langle n,m|), ~~~~~~~~~~ \\
    \delta \rho_b && = \frac{1}{2}\sum_{n,m} C_n^2 (\lambda_{m+1} (m+1) - m \lambda_m) |n,m\rangle \langle n,m|.
\end{eqnarray}
The Chernoff bound is again given by Eq. \eqref{Q_coherent} and in this case, $|i \rangle = |n,m \rangle$ and $\chi_i = C_n \sqrt{\lambda_m}$. The second subtracted term due to $\delta \rho_b$ reduces to 
\begin{equation}
   \frac{r^2}{4} \sum_{m = 0}^\infty \frac{[(m+1) \lambda_{m+1} - m \lambda_m]^2}{\lambda_m},
    \label{secondQTMSV}
\end{equation}
whose maximization condition is again independent of the state. The minimisation of $\mathcal{Q}$ depends only on the first subtracted term,
\begin{equation}
    \frac{\kappa (1 + 2 \bar{n})^2}{2 \bar{n}} \sum_{n = 0}^\infty (n+1) \mathcal{M}(C_n^2, \frac{\bar{n}}{1+\bar{n}}C_{n+1}^2),
    \label{firstQTMSV}
\end{equation}
where $\mathcal{M}(x,y) = \frac{4xy}{(\sqrt{x} + \sqrt{y})^2}$. It can be shown \cite{Optimalprobes} that Eq. \eqref{firstQTMSV} is maximised, subject to normalisation and energy constraints, by $C_n^2 = N_S^n/(1 + N_S)^{n+1}$ which represents the TMSV state. Therefore, the TMSV state minimises $\mathcal{Q}$, thereby giving the optimal two mode state for quantum illumination involving a light absorbing target. \hfill $\blacksquare$

\section{Conclusion}
\label{sec:conclu}

Modelling the target in a realistic manner is one of the most crucial tasks of the illumination process. In earlier works, targets were modelled using a beam splitter with a low reflectivity. However,  the beam splitter is seldom  perfect  and some amount of light gets inevitably lost or absorbed. A realistic model of the target should take into account these inherent target characteristics.  Towards fulfilling it, we introduced a target incorporating an additional feature of absorption on top of reflection and transmission which we designed via two auxiliary and a primary beam splitters, thereby ensuring it to be implemented easily in an optical setup. 

Our investigation revealed that the absorption of light by the target leads to an enhancement in the performance of the illumination protocol with the coherent state, thereby improving the bound in the classical domain and  the quantum illumination (QI) process involving a signal-idler setup via the two mode squeezed vacuum (TMSV) state. Unlike the non-absorbing target, we interestingly found that the Chernoff bound, the upper bound in the error probability of the detection, depends on the loss parameter and the mean photon number, thereby not reducing to the  Uhlmann's fidelity. 
We also proved that the coherent state for the single mode (classical) scheme  is optimal when both the reflectivity of the primary beam splitter and the absorption parameter are very low. 
It immediately prompted us to interpret the positive value in the difference between the Chernoff bounds for the coherent  and the TMSV states as the quantum advantage offered by the TMSV state.
We further reported that the quantum advantage decreases with the absorption parameter of the target, thereby furnishing the coherent state as the better probe in intermediate to high absorption regimes. Interestingly, for a light absorbing target, the background thermal noise enhances the detection efficiency with the increase of  the absorbing power of the target while  for the absorption-free targets, the thermal noise reduces the detection efficiency.
Our work widens the model for the target in the illumination procedure taking it a step forward to reality.

\section*{Acknowledgement} RG, SR, and ASD acknowledge the support from Interdisciplinary Cyber Physical Systems (ICPS) program of the Department of Science and Technology (DST), India, Grant No.: DST/ICPS/QuST/Theme- 1/2019/23. TD acknowledges support by Foundation for Polish Science (FNP), IRAP project ICTQT, contract no. 2018/MAB/5, co-financed by EU Smart Growth Operational Programme. 
	

\appendix	
\section{The transmission and absorption matrices}
\label{app:1}
Here we provide a sketch of the derivation of the transmission matrix $\mathrm{\hat T}$ and the absorption matrix $\mathrm{\hat A}$. First, note that the output modes of the two auxiliary beam splitters modelling the absorption, are related to the input modes, by 
\begin{eqnarray}
\hat a_{T'} &=& \sqrt{t} \hat a_T + \sqrt{r}  \hat g_2, \nonumber \\
\hat h_{2} &=& - \sqrt{r} \hat a_T + \sqrt{t}  \hat g_2,
\label{eq:absorbBS1}
\end{eqnarray}
and 
\begin{eqnarray}
\hat a_{S'} &=& \sqrt{t} \hat a_S + \sqrt{r}  \hat g_1, \nonumber \\
\hat h_{1} &=& - \sqrt{r} \hat a_S + \sqrt{t}  \hat g_1.
\label{eq:absorbBS2}
\end{eqnarray}
And for the final beam splitter, one has
\begin{eqnarray}
\hat b_{S} &=& \sqrt{\tau} \hat a_{T'} + \sqrt{\kappa}  \hat a_{S'}, \nonumber \\
\hat b_{T} &=& - \sqrt{\kappa} \hat a_{T'} + \sqrt{\tau}  \hat a_{S'}.
\label{eq:primaryBS}
\end{eqnarray}
 Note that the auxiliary beam splitters have a transmission coefficient $t$ while the primary one has $\tau$, as shown in  Fig. \ref{fig:lossy_model}. The corresponding reflectivities are denoted as $r$ and $\kappa$ respectively.
 Following the mode transformation in Eqs. \eqref{eq:absorbBS1} to \eqref{eq:primaryBS},  we obtain
 \begin{eqnarray}
\begin{pmatrix}
\hat{b}_S \\ \hat{b}_T 
		\end{pmatrix} = \mathrm{\hat T}\begin{pmatrix}
\hat{a}_S \\ \hat{a}_T
		\end{pmatrix} + \mathrm{\hat A}\begin{pmatrix}
\hat{g}_1 \\ \hat{g}_2
		\end{pmatrix},
 \end{eqnarray}
 where $\mathrm{\hat T} = \sqrt{t} {\hat U}$ and $\mathrm{\hat A} = \sqrt{r} {\hat U}$, with
 \begin{eqnarray}
&&\mathrm{\hat U} = \begin{pmatrix}
\sqrt{\kappa} & \sqrt{\tau } \\ \sqrt{\tau } & -\sqrt{ \kappa}
		\end{pmatrix}.
 \end{eqnarray}
It trivially follows from the above analysis that
\begin{eqnarray}
\mathrm{\hat T}^{\dagger}\mathrm{\hat T} = t ~\mathbb{I}, \text{ and } \mathrm{\hat A}^{\dagger}\mathrm{\hat A} = (1-t) ~\mathbb{I},
\end{eqnarray} 
establishing $1-t = r$ as the loss parameter.
 
\section{Computation of $\rho_1$}
\label{app:2}
We now present the computation of $\rho_1$, which corresponds to the case where the target is present by using the coherent and the TMSV state as the probe states. Recall that we can group the canonical position and momentum operators of an $\mathcal{N}$ mode Gaussian state in the vector as
\begin{equation}
\boldsymbol{\hat{R}} = (\hat{x}_1,\hat{p}_1,...,\hat{x}_\mathcal{N},\hat{p}_\mathcal{N})^T
\end{equation}
and the Gaussian state is completely specified by the displacement vector and the covariance matrix, given by
\begin{eqnarray}
&&d_i = \langle \hat{R_i} \rangle, \\
&&\sigma_{ij} = \frac{1}{2} [\langle \hat{R}_i \hat{R}_j + \hat{R}_j \hat{R}_i \rangle - 2 \langle \hat{R}_i \rangle \langle \hat{R}_j \rangle],
\end{eqnarray}
where $i,j = 1,2,...,d$ with the  annihilation operator being $\hat{a} = (\hat{x} + i \hat{p})/2$ in terms of the mode quadratures.

\subsection{Coherent states}
When the coherent states are used as  probes for target detection, we lay out the sketch of the derivation for $\rho_1$. Since the initial input states are all Gaussian and as mentioned earlier, the optical setup modelling loss preserves Gaussianity, one can efficiently compute $\rho_1$ in the phase space. \\
As clearly illustrated in the schematic diagram,  there are four input modes, two of which correspond to the thermal and the coherent state, while the remaining two modes corresponds to the ground states of the device, which are the vacuum states. Note that the combined four states are Gaussian and are described by the following displacement vector and covariance matrix:
\begin{eqnarray}
\boldsymbol{d} &=& d_T \oplus d_{g_1g_2} \oplus d_{CS} \nonumber \\
&=& (0,0)^T \oplus (0,0,0,0)^T \oplus (2\sqrt{N_S},0)^T, \nonumber \\
\boldsymbol{\sigma} &=& \sigma_T \oplus \sigma_{g_1g_2} \oplus \sigma_{CS} = (2\bar{n} + 1) \mathbf{I_2} \oplus \mathbf{I_4} \oplus \mathbf{I_2},
\end{eqnarray}
where $N_S$ is the signal strength of the coherent state while $\bar{n}$ denotes the average number of photons in the thermal state with $\mathbf{I}_4 = \text{diag}\{1,1,1,1\}$. The joint operation of the two auxiliary beam splitters is described by the  symplectic operation, 
\begin{eqnarray}
\boldsymbol{S_A} = S_1 \oplus S_2, 
\label{eq:sym1}
\end{eqnarray}
where
\begin{eqnarray}
S_{1(2)}(r) = \begin{pmatrix}
\sqrt{1-r} & 0 & \sqrt{r} & 0 \\
0 & \sqrt{1-r} & 0 & \sqrt{r} \\
-\sqrt{r} & 0 & \sqrt{1-r} & 0 \\
0 & -\sqrt{r} & 0 & \sqrt{1-r} 
\end{pmatrix},  
\label{eq:bssym1}
\end{eqnarray}
with $t$ being the transmission coefficient of the auxiliary beam splitters and follows the  same mode transformations as given in Eqs. \eqref{eq:absorbBS1} -- \eqref{eq:primaryBS}. The resultant state is now described by these updated moments as
\begin{eqnarray}
\boldsymbol{d} &\rightarrow& \boldsymbol{S_A}\boldsymbol{d}, \nonumber \\
\boldsymbol{\sigma} &\rightarrow& \boldsymbol{S_A}\boldsymbol{\sigma}\boldsymbol{S_A^T}.
\label{eq:update1}
\end{eqnarray}
Tracing out the device modes, we now have the effective state that impinges in the primary beam splitter,  described by
\begin{eqnarray}
\boldsymbol{d_{es}} &=& (0,0,0,2\sqrt{t N_S})^T, \nonumber \\
\boldsymbol{\sigma_{es}} &=& (1 + 2\bar{n}t) \mathbf{I_2} \oplus \mathbf{I_2}.
\end{eqnarray}
The action of the primary beam splitter with transmissivity $\tau$ is described by the symplectic operation,
\begin{eqnarray}
\boldsymbol{S_P} = S_1(\tau),
\end{eqnarray}
where the form of $S_1$ is given in Eq. \eqref{eq:bssym1}. The state can be updated by following the same rules as in Eq. \eqref{eq:update1}. Finally,  the state that impinges on the detector, $\rho_1$  described by the moments reads as
\begin{eqnarray}
\boldsymbol{d_1} &=& (2\sqrt{t \kappa N_S}, 0)^T, \nonumber \\
\boldsymbol{\sigma_1} &=& (1 + 2\bar{n} t\tau )   \mathbf{I_2}.
\end{eqnarray}

\subsection{TMSV state} 
 Let us compute the displacement vector and the covariance matrix of the state that goes to the detector, $\rho_1$, when the TMSV state is used for target detection whose one part is used as the probe, while the other part forms the idler that directly goes to the detector. A TMSV state of squeezing strength $r$ is described by 
\begin{eqnarray}
\boldsymbol{d}_{TMSV} &=& (0,0,0,0)^T, \nonumber \\
\boldsymbol{\sigma}_{TMSV} &=& \begin{pmatrix}
(2N_S + 1) ~\mathbf{I_2} & 2\sqrt{N_S(1+ N_S)} ~\mathbf{Z_2} \\
2\sqrt{N_S(1+ N_S)} ~\mathbf{Z_2} & (2N_S + 1) ~\mathbf{I_2}
\end{pmatrix}. \nonumber \\
\label{eq:tmsvcov}
\end{eqnarray}
The displacement vector and covariance matrix of  the entire state that impinges on the auxiliary beam splitters are given by
\begin{eqnarray}
\boldsymbol{d} &=& d_T \oplus d_{g_1g_2} \oplus d_{TMSV} = (0,0,0,0,0,0,0,0,0,0)^T, \nonumber \\
\boldsymbol{\sigma} &=& \sigma_T \oplus \sigma_{g_1g_2} \oplus \sigma_{TMSV} \nonumber \\ &=& (2\bar{n} + 1) \mathbf{I_2}  \oplus \mathbf{I_4} \oplus \sigma_{TMSV},
\end{eqnarray}
where $\bar{n}$ is the average number of photons and the form of $\sigma_{TMSV}$ is given in Eq. \eqref{eq:tmsvcov}. In this case, the action of the two auxiliary beam splitters on all the modes are given by the following symplectic operation:
\begin{eqnarray}
\boldsymbol{S_A'} = \boldsymbol{S_A}\oplus \mathbf{I_2}, 
\end{eqnarray}
where $\boldsymbol{S_A}$ is given in Eq. \eqref{eq:sym1}. The updated state is obtained by applying the rules given in Eq. \eqref{eq:update1}. After tracing out the device modes, we end up with the effective state that impinges on the primary beam splitter, given by
\begin{eqnarray}
\boldsymbol{d_{es}} &=& (0,0,0,0,0,0)^T, \nonumber \\
\boldsymbol{\sigma_{es}} &=& (1+ 2\bar{n}t)\mathbf{I_2} \oplus \sigma',
\end{eqnarray}
where
\begin{eqnarray}
\sigma' = \begin{pmatrix}
(1 + 2tN_S)~\mathbf{I_2} & 2\sqrt{t N_S(1+N_S)}  ~\mathbf{Z_2} \\
2\sqrt{tN_S(1+ N_S)}  ~\mathbf{Z_2} & (1 + 2N_S) ~\mathbf{I_2}
\end{pmatrix}. \nonumber \\
\end{eqnarray}
The effective state  falls on the primary beam splitter  whose action is described by the  symplectic operation as
\begin{eqnarray}
\boldsymbol{S_P} = S_1(\tau) \oplus \mathbf{I_2},
\end{eqnarray}
where the form of $S_1$ is given in Eq. \eqref{eq:bssym1}. The final output state is obtained via the Eq. \eqref{eq:update1}. Finally, tracing out the transmitted part of the signal, we arrive at the phase space description of $\rho_1$ when the TMSV state is used as the probe:
\begin{equation}
\boldsymbol{d_1} = (0, 0, 0, 0)^T,
\end{equation}
\begin{equation} \boldsymbol{\sigma_1} = \begin{pmatrix}
\big(1 + 2t(\bar{n}\tau  + \kappa N_S)\big)~\mathbf{I_2} & 2\sqrt{t\kappa N_S(1+ N_S)}  ~\mathbf{Z_2} \\
2\sqrt{t\kappa N_S(1+ N_S)}  ~\mathbf{Z_2} & (2N_S + 1) ~\mathbf{I_2} 
\end{pmatrix}. 
\end{equation}

 	\bibliography{bib}
 	\bibliographystyle{apsrev4-1}
	
\end{document}